\documentclass[11pt]{article}

\input epsf.sty

\usepackage{graphicx,amsmath,amssymb}
\def\lromn#1{\uppercase\expandafter{\romannumeral#1}}

\begin{document}

\begin{flushright}
%\today \\
\end{flushright}

\begin{center}
\begin{Large}
{
Magnetic skyrmions as host of neutrino mass spectroscopy
}
\end{Large}

\vspace{1cm}
M.~Yoshimura

Research Institute for Interdisciplinary Science,
 Okayama University 
Tsushima-naka 3-1-1 Kita-ku Okayama
700-8530 Japan

\end{center}

\vspace{5cm}

\begin{center}
\begin{Large}
{\bf ABSTRACT}
\end{Large}
\end{center}

Large magneto-electric effect in multi-ferroics and topological protection against decays of
excited magnetic skyrmions 
are an ideal setting for detecting radiative emission of neutrino pair (RENP). 
We propose a new RENP scheme using multi-ferroic magnetic skyrmion condensates
 in order to determine the absolute neutrino mass
including the mass hierarchy patterns,
the type of neutrino mass (Majorana vs Dirac), one of CP violation parameters, and
to detect relic 1.9 K cosmic neutrino.
We critically examine the possibility of feasible experiments,
and show what are required as good targets for neutrino physics.

\vspace{5cm}

\newpage

{\bf Introduction}
\hspace{0.3cm}
The discovery of neutrino oscillation \cite{pdg} has established that at least two of three
neutrinos have finite masses with their mass eigenstates mixed
in weak decay of nuclei and elementary particles.
Combined with the cosmological microwave fluctuation and simulation of structure formation \cite{pdg},
the heaviest neutrino has a mass in the range $m_3 = 60 \sim 200 $ meV,
the next heaviest of  $ m_2 =\sqrt{m_3^2 - (50 {\rm meV})^2} $,
and the lightest of $m_1 = \sqrt{m_2^2 - (10 {\rm meV})^2}$ 
for the normal  hierarchy (NH) mass pattern.
For the inverted hierarchy (IH) mass pattern two measured mass difference values, 
$(50 {\rm meV})^2$ and $ (10 {\rm meV})^2$, are interchanged in the formula.
Neutrino oscillation experiments are sensitive  only to mass squared differences
and they left important questions unanswered:
(1) absolute value of neutrino masses,
(2) nature of neutrino masses, either of Dirac type or of Majorana type,
(3) CP violation (CPV) parameters (one phase for Dirac and three phases for Majorana).
There are  on-going experiments \cite{kamland}, \cite{katrin}, \cite{pdg} 
that attempt to partially answer questions,
(1) and (2), using nuclear targets.
So far results of time consuming  efforts have yielded only limits
on neutrino parameters.
One of the major goals in the next round of oscillation experiments is to measure
one of CPV phases.
One may add  to this list (4) detection of cosmic relic neutrino of 1.9 K \cite{yst relic}
which clarifies the state of universe at one second after the big bang.

Atomic targets are better suited than nuclear targets to answer these questions, 
since available energy differences $< O(10)$eV are closer
 to neutrino masses. But detection rate when events occur incoherently
are much less at unobservable levels.
It was recently suggested  \cite{my-prd-07}, \cite{renp overview}    that a prepared
macro-coherence of states  enhances rates to circumvent the smallness of atomic rates
in the  de-excitation process
$|e \rangle \rightarrow |g \rangle + \gamma + \nu_i \bar{\nu}_j$;
 radiative emission of neutrino pair (RENP) with $\nu_i, \bar{\nu}_j\,, i,j =1,2,3$ mass eigenstates. 
The target phase coherence between an excited state $|e\rangle $ and the ground state $|g\rangle $
may be realized by good-quality excitation laser.
Experiments \cite{psr exp} in QED second order process of para-hydrogen, two-photon emission
from a vibrationally excited state,  
confirmed  that the coherence close to necessary levels is 
achieved, with rate enhancement
of order $10^{18}$.
The project of macro-coherent RENP was termed neutrino mass spectroscopy 
\cite{renp overview}, \cite{yst relic},
since it covers all four  raised questions, (1) $\sim$ (4).
Note that one does not need any specific theory of neutrino mass 
beyond the standard theory of particle physics except introduction
of the neutrino mass term, 
in order to advance this project.
There is however a serious problem envisaged in RENP schemes using a gas due to
small available target atom numbers, hence it is desirable to
use solid environments for  detection. But
various de-coherence processes are expected to be very fast in solids in conflict with the
macro-coherence.

In the present work we propose to solve the de-coherence problem in solids by
a topology protection.
A series of recent experiments in condensed matter physics  discovered
topologically stable magnetic skyrmions in a variety of magnetic systems \cite{skyrmion review}.
Magnetic skyrmions are characterized by integer $Z$ topological charges,
$ \int d^2 r\, \vec{n}\cdot \partial_x \vec{n} \times \partial_y \vec{n}/ 4\pi$
with $\vec{n} = \vec{S}/S$ the unit spin vector, 
as defined by mapping of a sphere onto two dimensional plane.
The topology protection realizes a long relaxation time, which is proposed below to use
 in applications to fundamental physics, which may be called topological RENP experiment.
Calculated event rates are in proportion to squared magneto-electric strength, hence
insulating magnetic skyrmions are candidate hosts.

We use in the present work the natural  unit such that $\hbar = c= k_B = 1$ to simplify theoretical  formulas.
In this unit 1 eV $= 1/( 1.97 \times 10^{-5}{\rm cm})  =  3/(1.97 \times 10^{-15}{\rm sec}) $
and 1 K $=0.086$ meV.

\vspace{0.5cm}
{\bf Multi-ferroic coupling in  RENP}
\hspace{0.3cm}
Neutrino pair emission from atomic electron occurs with a hamiltonian density, 
$\nu_i ^{\dagger} \vec{\sigma} \nu_j \cdot e^{\dagger} \vec{\sigma} e$,
 neutrino-pair spin  times electron spin $\vec{S}= \vec{\sigma}/2$,
written in terms of  field operators  \cite{renp overview}.
This four-Fermi interaction consists of added W- and Z-exchange contributions.
Mass eigenstate fields $\nu_i$'s 
are related to the electron neutrino $\nu_e $ 
that appears in beta decay by $\nu_e = \sum_i U_{ei}\nu_i \,,$
where $ (U_{\alpha i}) \,, \alpha = e,\mu,\tau,$ is the $3\times 3$ unitary mixing matrix.
Relevance of the electron spin in the hamiltonian
 suggests how the magnetic order in solids may help in RENP.
Since it is practically impossible to detect neutrinos in atomic experiments,
one needs detection of photons to probe neutrino properties. 
For this  the weak interaction is connected to QED
electric dipole interaction, $e \vec{E}\cdot e^{\dagger} \vec{r} e $
in the second order perturbation theory, 
to give a product of spin and dipole, $ \vec{S}_i \, e \vec{r}_j  $
for electronic transition.
The RENP probability amplitude is then given by
\begin{eqnarray}
&&
\frac{2 G_F}{\sqrt{2}}\, a_{ij} \,\nu_i ^{\dagger} \vec{\sigma} \nu_j \cdot
\langle g| \sum_n \left( \frac{\vec{S} | n\rangle \langle n |\vec{d}  }
{\epsilon_e - \epsilon_n - \omega} + 
 \frac{\vec{d} | n\rangle \langle n |\vec{S}  }
{\epsilon_g - \epsilon_n + \omega} 
\right)| e\rangle \cdot \vec{E}_{\gamma}
\,, \hspace{0.5cm}
a_{ij} = U_{ei}^*U_{ej} - \frac{1}{2}\delta_{ij}
\,,
\label {renp amplitude}
\end{eqnarray}
(with $G_F = 1.17 \times 10^{-5}{\rm GeV}^{-2}$ the Fermi coupling constant)
using the energy conservation $\epsilon_g  + \omega = \epsilon_e - E_i - E_j$
to eliminate the neutrino pair energy $ E_i + E_j\,, E_i = \sqrt{p_i^2 + m_i^2}$ 
in favor of the photon energy $\omega$.
We shall give an example of states, $|e \rangle, |n \rangle, |g\rangle$ later.

The summed bracket quantity  in eq.(\ref{renp amplitude}) may be
rewritten by introducing an average intermediate state energy $\bar{\epsilon}_n$
and by multiplying 
the magnetic moment $g \mu_B$ times  the squared number density $N^2$ of
excited atoms, to give
\begin{eqnarray}
&&
N^2 (\frac{2m_e} {g })^2\sum_n  \left( \frac{\vec{\mu}_i | n\rangle \langle n |\vec{d}_j  }
{\epsilon_e - \epsilon_n - \omega} + \frac{\vec{d}_j  | n\rangle \langle n | \vec{\mu}_i }
{\epsilon_g - \epsilon_n + \omega}
\right) \equiv h(\omega) \vec{M}_i \vec{P}_j
\,, \hspace{0.5cm}
h(\omega) = 
\frac{\epsilon_{eg} - 2 \bar{\epsilon}_n} 
{(\omega - \bar{\epsilon}_n) (\epsilon_{eg} - \bar{\epsilon}_n - \omega  )} 
\,,
\label {energy denom}
\end{eqnarray}
where $\vec{M}, \vec{P}$ are magnetization and electric polarization
of target medium, respectively,
assumed to be uniform in crystals.

A large  product $MP$ may emerge  in multi-ferroic materials,
objects in which both ferro-electric and ferro(or anti-ferro)-magnetic orders
co-exist \cite{multiferroic}, \cite{tokura et al}.
Calculated event rates of RENP in multi-ferroics  depend on an overall factor given by
$G_F^2 ( 2m_e MP/N)^2 V^2 $ ($V$ a target volume) times a power of
atomic level spacing between $|e\rangle $ and $|g\rangle $, $\epsilon_{eg}^7$.
Taking typical values for $M, P, N, \epsilon_{eg}$, along with the phase space
volume of neutrino pair, gives  total RENP rates of order,
\begin{eqnarray}
&&
\frac{6 }{ (2\pi)^4}
G_F^2 (\frac{ 2 m_e MP V} {g  N})^2 \epsilon_{eg}^7 
\nonumber 
\\ &&
\sim 1.0 \times 10^{-5}  {\rm sec}^{-1}\,
(\frac{V}{  {\rm cm}^3})^2 (\frac{(5 \times 10^{-8}{\rm cm})^{-3} }{N })^2 
(\frac{M }{g \mu_B ( 5 \times 10^{-8}{\rm cm})^{-3}})^2 (\frac{P }{ \mu {\rm C}/{\rm m}^2})^2
(\frac{\epsilon_{eg} } {{\rm eV}})^7
\,.
\end{eqnarray}

\vspace{0.5cm}
{\bf RENP spectrum}
\hspace{0.3cm}
There are six distinct pair thresholds denoted by $(ij)$ at $\omega_{ij} = \epsilon_{eg} - m_i - m_j$  
\cite{no momentum conservation},  rate rises  being determined by
neutrino mass mixing parameters, including CP violating phases.
Distinction of Majorana from Dirac neutrinos  becomes possible
by exploiting the very nature of Majorana particle being its own anti-particle;
$\nu_i = \bar{\nu}_i $.
Theoretical prediction  of Majorana-pair production is 
governed by anti-symmetric wave functions of two identical fermions, which
gives spectra different from  Dirac-pair of distinguishable neutrino and anti-neutrino
\cite{my-prd-07}.

For spectrum calculation
recall that the neutrino  helicity  and the
momentum ($\vec{p}_i$) are summed due to their
detection being impossible, to give the neutrino phase space integral,
including the Pauli blocking of relic neutrinos,
\begin{eqnarray}
&&
\hspace*{-0.5cm}
\int \frac{d^3p_i d^3p_j}{(2\pi)^5} \delta(\epsilon_{eg} - \omega - E_i - E_j)
(1  - \delta_M \frac{m_i m_j}{E_i E_j} )
 = \frac{ 4}{(2\pi)^3} 
\epsilon_{eg}^5 \left(
G(\frac{\omega}{\epsilon_{eg}}, \frac{m_i}{\epsilon_{eg}}, \frac{m_j}{\epsilon_{eg}})
- \delta_M H (\frac{\omega}{\epsilon_{eg}}, \frac{m_i}{\epsilon_{eg}}, \frac{m_j}{\epsilon_{eg}})
\right)
\,, 
\label {md distinction}
\\ &&
G(x, a, b) =  \int_{a}^{1- x -b} dy
\left( (y^2 - a^2)(\,(1-x-y)^2 - b^2) \right)^{1/2} y (1-x - y) 
(1 - f_i\,) (1 - f_j\,)
\,,
\\ &&
H(x, a, b) =  a b \int_{a}^{1- x -b} dy
\left( (y^2 - a^2)(\,(1-x-y)^2 - b^2) \right)^{1/2} 
(1 - f_i\,) (1 - f_j\,)
\,,
\end{eqnarray}
where $f_i =  1/(e^{p_i/T_0} +1)$ is the Fermi-Dirac  distribution function
of $p_i = \sqrt{E_i^2 - m_i^2}$, expressed in terms of integration variables $x,y,a,b$,
with $T_0 = 1.9\,$K the relic neutrino temperature  \cite{yst relic}.
The Majorana/Dirac distinction appears in the $\delta_M$ term of eq.(\ref{md distinction}): 
$\delta_M = 1$ for Majorana neutrinos and $=0$ for Dirac neutrinos.

The differential spectrum rate is given by
\begin{eqnarray}
&&
\frac{d^2 \Gamma}{d\omega d\Omega} = \frac{ 6}{  (2\pi)^4} G_F^2
( \frac{2m_e }{ g})^2  ( \frac{MP V}{ N})^2 \sin^2 \theta \cos^2 \varphi \,
 \epsilon_{eg}^5
F(\omega; m_i)
\,,
\label {renp spectrum rate}
\\ &&
F(\omega; m_i) =
\omega^3\, |h(\omega) |^2
\sum_{ij} \left(
|a_{ij} |^2 
G(\frac{\omega}{\epsilon_{eg}}, \frac{m_i}{\epsilon_{eg}}, \frac{m_j}{\epsilon_{eg}})
- \delta_M \Re (a_{ij}^2 )
H(\frac{\omega}{\epsilon_{eg}}, \frac{m_i}{\epsilon_{eg}}, \frac{m_j}{\epsilon_{eg}})
\right)
\,.
\label {renp spectrum shape}
\end{eqnarray}
$\theta$ is the  photon emission angle measured from
$\vec{P}-$direction,
and $\varphi$ is the photon linear polarization ($\vec{e} $) angle from the plane made of
$\vec{P}$ and photon wave vector such that $\vec{P}\cdot \vec{e} = P\sin \theta \cos \varphi$. 
Note that except the average value $\bar{\epsilon}_n$ in $h(\omega)$
of eq.(\ref{energy denom}) and $\epsilon_{eg}$,
RENP rate formula contains measured quantities of target atoms such as $M, P, N$.
The massless limit of neutrinos without the relic effect is easily worked out, to give the integral,
$\epsilon_{eg}^5 \,G(\omega/\epsilon_{eg} ,0,0)\sum_{ij}|a_{ij} |^2 
= ( \epsilon_{eg} - \omega)^5/40$.
The global spectrum shape is of the form,
$\omega^3 |h(\omega) |^2 ( \epsilon_{eg} - \omega)^5 d\omega $,
which is modified by finite neutrino mass function
$F(\omega; m_i \neq 0)$.

\begin{figure}[htbp]
 \centering
\includegraphics[height=5cm]{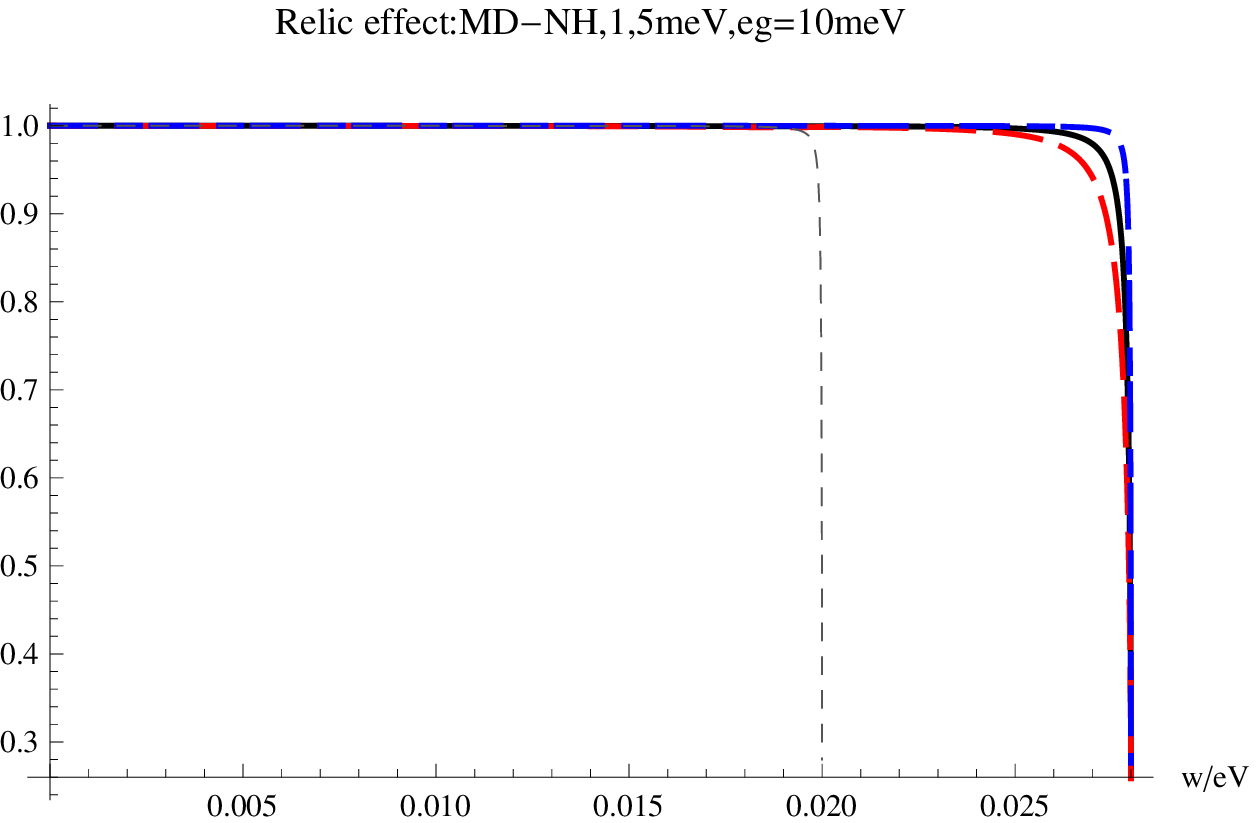} 
 \vspace{0.5cm}
  \includegraphics[height=5cm]{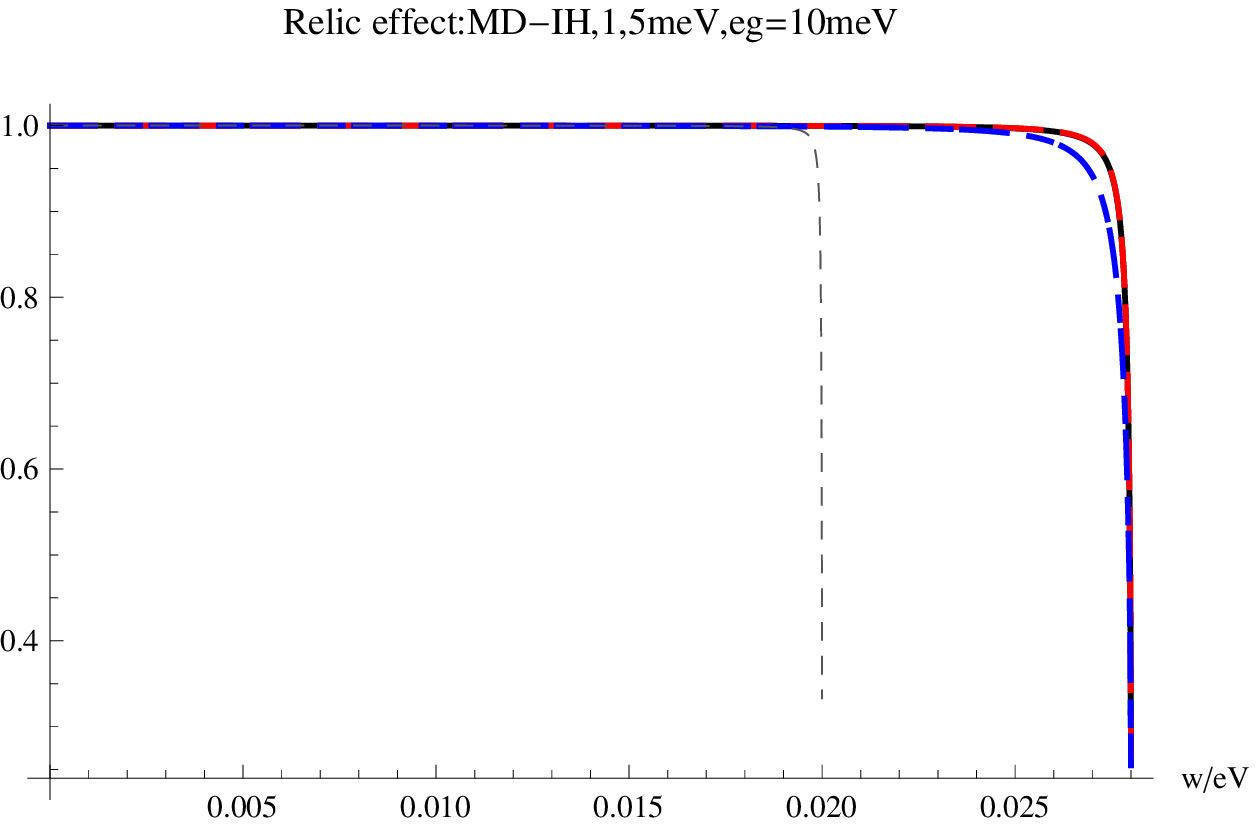}
   \caption{
Theoretically calculated  ratio of two rates: with the relic Pauli blocking and without it:
$\epsilon_{eg} = 30$ meV with $\bar{\epsilon}_{n} = 40$meV in $f(\omega)$ of eq.(\ref {energy denom}),
all assuming CP conservation and Majorana NH neutrinos.
Assumed parameters are  the smallest mass 1 meV, 1.9 K 
in solid black,  5 meV, 1.9 K in thin dashed black, 1 meV, 1.9 K x 1.5
in dashed red, 1 meV, 1.9 K x 0.5  in dash-dotted blue.
Left panel for NH and right panel for IH.
}
   \label {relic effect}
\end{figure}

Before we discuss a specific target example of spectrum shapes,
we demonstrate importance of  relic neutrino effects in Fig(\ref{relic effect}).
The ratio of two spectrum rates \cite{relic importance}, 
rate including the Pauli blocking  $1-f_i$  to  rate
without it, is plotted for $\epsilon_{eg} = $ 30 meV.
Actual experimental data cannot disentangle relic neutrino effect of 1.9 K, hence
this ratio may effectively be regarded as experimental values divided by a hypothetical result calculated
without the relic neutrino.
Mass determination to sensitivity $> O(1)$ meV and
the Majorana/Dirac distinction is not difficult 
for $\epsilon_{eg}=$ 30 meV.
We find it remarkable that the topological RENP greatly improves the sensitivity
of neutrino mass spectroscopy including the relic neutrino detection.
In the rest of presentation we include the important effect of 1.9 K relic neutrino \cite{cmb}.

\vspace{0.5cm}
{\bf An example of  RENP scheme in multi-ferroic skyrmion lattice}
\hspace{0.3cm}
Most magnetic skyrmions appear in metals, but 
recently, insulating multi-ferroic skyrmion lattice (SkL)  has been discovered in Cu$_2$OSeO$_3$ as its host 
\cite{multi-f skyrmion}, \cite{msk-multiferroic 2}, \cite{skyrmion review}.
Calculations in \cite{multi-f skyrmion 3}, \cite{multi-f skyrmion 4}
indicate that both parallel and perpendicular $\vec{M}\,, \vec{P}$'s of
 large values  may arise consistently with
experimental results.
Measured values of this SKL are
$P_0 =  0.2 \mu {C}/ {\rm m}^2
\,, \;
M_0 =  0.1 \mu_B/ {\rm Cu}^{2+}
\,, \;
N = 1/( 8.9 \times 10^{-8} {\rm cm})^{-3}
\,.$
The unit cell of Cu$_2$OSeO$_3$ is made of four bypyramids of altogether 16 Cu$^{2+}$ magnetic ions.
According to density functional theory (DFT) calculation \cite{multi-f skyrmion 4}
the ground state  consists of four spin triplets $| 1 \rangle$ of 4 Cu$^{2+}$.
When one of bypyramids is excited to either of
two singlets $| 0 \rangle_i, i = 1,2$ of energy $\sim 24\,$meV, 
a quintet $| 2 \rangle$ of energy $\sim 29\,$meV, or two triplets $| 1' \rangle_i, i = 1,2$
of energy $\sim 39\,$meV,
they form the first groups of excited skyrmions. 
In RENP rate estimate we take as relevant states,
$|e \rangle = | 2 \rangle\,, | n \rangle = | 1' \rangle_i\,, | g\rangle = | 1 \rangle$
in eq.(\ref{energy denom}).

We illustrate  numerical results of calculated RENP spectral shapes
in Fig(\ref{md distinction, cpv}).
Due to the closeness of this system ($\epsilon_{eg} = 29\,$meV) to neutrino masses in the 10 meV range,
there are  good sensitivities both to Majorana/Dirac distinction and to
one of Majorana CPV phases.
For determination of neutrino properties and parameters
 the spectrum shape  $ F(\omega; m_i)$ of eq.(\ref{renp spectrum shape})
is all one needs.
But the feasibility of RENP does depend on absolute rate values.
In order to obtain absolute rates, one should multiply numbers in Fig(\ref{md distinction, cpv})
by $4 \times 10^{-9}$sec$^{-1}$/eV.
Hence the example of Cu$_2$OSeO$_3$ gives rates too small to be detectable.
Possible improvements shall be discussed later.

\begin{figure}[htbp]
 \centering
  \includegraphics[height=5cm]{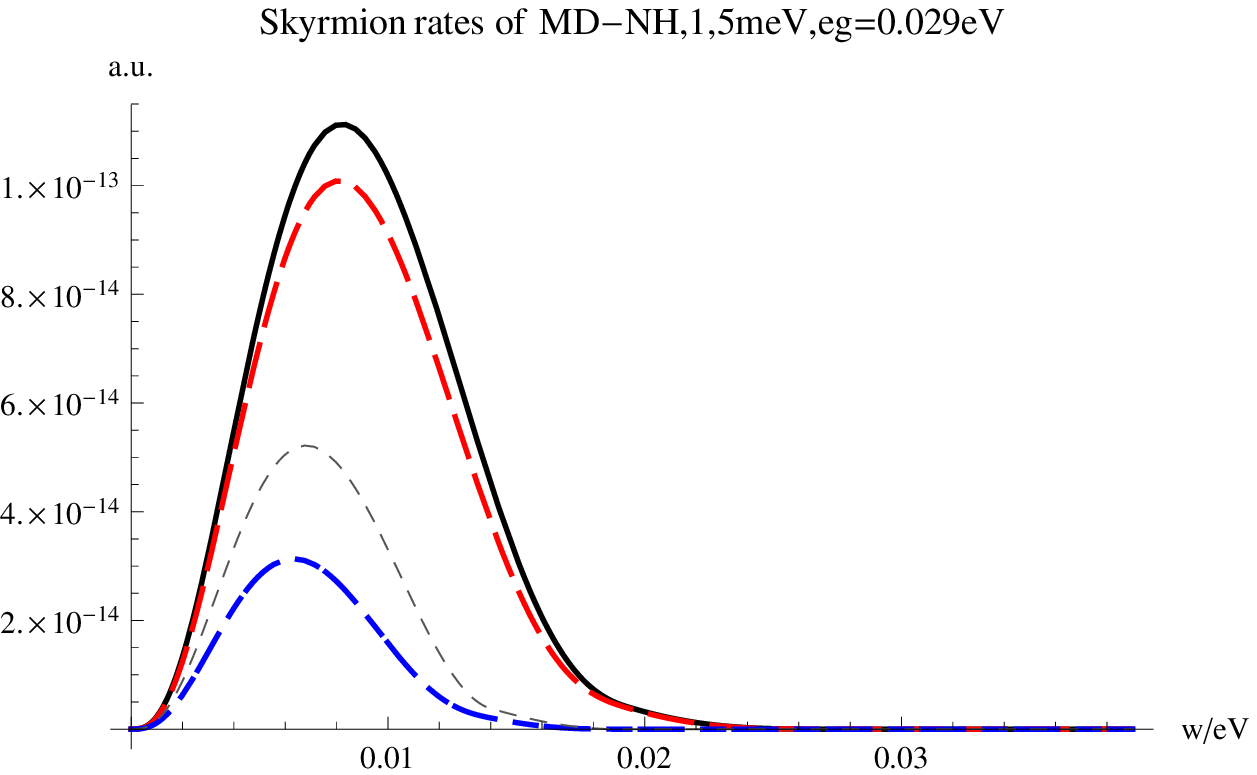}%
  \vspace{0.5cm}
  \includegraphics[height=5cm]{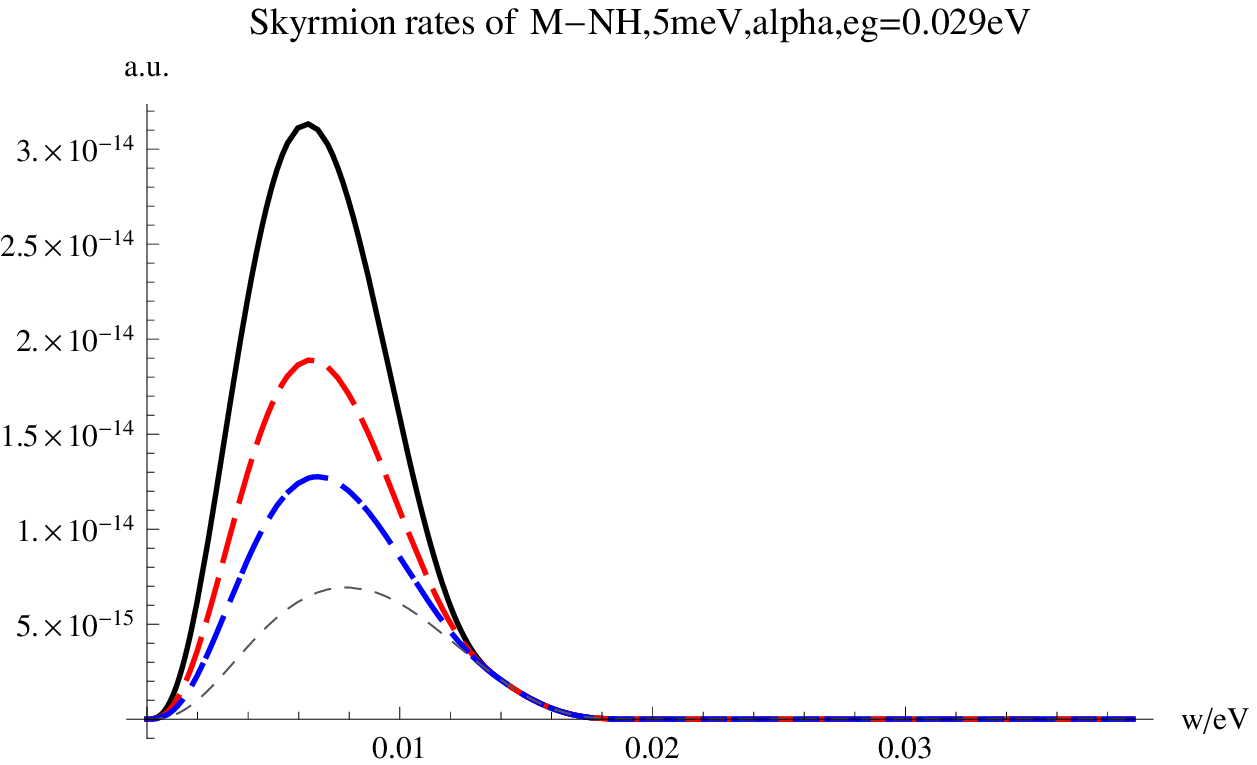}%
   \caption{Left panel. Spectrum sensitivity to Majorana and Dirac neutrino masses in NH case.
Dirac mass 1 meV in solid black, 5 meV in dotted black,
Majorana mass 1 meV in dashed red, and 5 meV in dash-dotted blue.
All cases assume CP conservation.
Right panel.
Spectrum sensitivity to Majorana CPV phases.
NH Majorana  of the smallest neutrino mass 1  meV is assumed:
$(\alpha, \beta, \delta) = (0,0,0)$ in solid black,
$ (\pi/8,0,0)$ in dashed red, $ (\pi/6,0,0)$ in dash-dotted blue,
and $ (\pi/4,0,0)$ in dotted black.
}
 \label{md distinction, cpv}
\end{figure}

Rate increase with the level spacing $\epsilon_{eg}$ is dramatic:
the total rate increases with a high power of $\epsilon_{eg}$  $\propto \epsilon_{eg}^7$
if $\epsilon_{ne} < \epsilon_{eg}$, as numerically confirmed. 
Hence it is critically important to find more insulating
skyrmions of larger energy gaps  to achieve larger rates.
Or one should find more appropriate excited levels in   Cu$_2$OSeO$_3$,
for instance 150 meV or a slightly higher.

\vspace{0.5cm}
{\bf Role of topology}
\hspace{0.3cm}
We assume that both the excited initial lattice state and the final one in RENP
are topologically stable with skyrmion condensates.
Difference (including the vanishing one) of 
topological quantum numbers is transfered to those of emitted particles,
since the overlap of wave functions of emitted particles and emitting
atoms is very small.
This becomes possible  if  three unit vectors $\vec{n}_i$ or RENP particles 
carry a topological integer value of 
$\int d^2 r\, \vec{n}_1 \cdot \partial_x \vec{n}_2 \times \partial_y \vec{n}_3/4\pi$.
Three spin vectors $\vec{n}_i$  can be  in the same plane:
the neutrino pair is emitted perpendicular to each other with
their spins almost (exactly for the massless case) along emitted directions 
and photon polarization out of the emission plane, corresponding to $\theta = \pi/2, \varphi=0$
in the rate formula.
This restriction gives a reduction of neutrino phase space integration by some amount,
but within a tolerable range of reduction.

The topological conservation law, on the other hand,
restricts possible modes and configurations of QED backgrounds.
The major QED background is three photon emission, with
polarization constraints similar to the RENP case.
Although the QED background has much larger rates than RENP,
specified characteristics of background events helps much
to reject backgrounds from the RENP signal.

\vspace{0.5cm}
{\bf Comments and prospects}
\hspace{0.3cm}
In the rate estimate given above, we took  modest $P,M, N$ values  of target Cu$_2$OSeO$_3$.
There are a few rooms for improvements if we use  better targets (yet to be identified):
(1) larger gaps, for instance $\epsilon_{eg} =2.9\,$eV $=$ 100 times Cu$_2$OSeO$_3$ value
gives $10^{14}$ larger rate,
(2) increase of  $10^3$ larger $P$ to $200 \mu $C/m$^2$ gives $10^6$ larger rate,
(3) decrease of spin density $N$ by 10 gives $10^2$.
There exist multi-ferroics without skyrmion structure satisfying requirements, (2) and (3).
Taken altogether, (1) $\sim$ (3) increase the rate by $10^{22}$, presumably in  the feasibility range
of order 0.5 sec$^{-1}$.

On theoretical fronts there are a number of works to be done:
(1)  DFT calculation of excited skyrmions to search for a level spacing $\epsilon_{eg} =$
a few times 100 meV (the best sensitivity to neutrino
mass determination expected),
(2)  detailed study of QED background rejection including effects of topology protection.

\vspace{0.5cm}
In summary,
multi-ferroic skyrmions provide the unique opportunity of unraveling the conundrum
of neutrinos deeply related to physics beyond the standard theory and fundamental problems of
cosmology.

%\begin{acknowledgments}
\vspace{1cm}
{\bf Acknowledgments}

Discussions on magnetic skyrmions with J. Akimitsu and N. Yokoi
are greatly appreciated.
This work is supported in part by JSPS KAKENHI Grant Numbers JP 
JP17H02895.
%\end{acknowledgments}


\begin{thebibliography}{99}
\bibitem{no momentum conservation}
In the topological RENP scheme here no momentum conservation among
three emitted particles, $\gamma, \nu_1, \nu_2$, is not imposed,
which gives  spectrum shapes given by eq.(\ref{renp spectrum rate}),
 different from other RENP schemes of \cite{renp overview}.
Contributions to RENP rates with the momentum conservation taken into account
is found to be much smaller than rates given here.
\bibitem{relic importance}
The reason the Pauli blocking distortion appears in a wider energy region of Fig(\ref{relic effect})
than the case of \cite{yst relic} is that
the phase space restriction without the momentum conservation \cite{no momentum conservation}
is much less severe.
The  Fig(\ref{relic effect}) type of plot makes it possible 
to use  the relic neutrino effect for identification of neutrino mass thresholds.
\bibitem{cmb}
Strictly, one should include the stimulated cosmic microwave background (CMB) emission
effect by multiplying  $1+f_{{\rm cmb}}(\omega)\,, 
f_{{\rm cmb}}(\omega) = 1/ (e^{\omega/T_{{\rm cmb}}} - 1)\,, T_{{\rm cmb}} \sim 3$K 
to the rate formula, eq.(\ref{renp spectrum rate}).
But this has effects only at $\omega <  T_{{\rm cmb}}$ where
rates are very small.
Hence we ignore this factor.
\end{thebibliography}
\end{document}